# An empirical study into the relationship between class features and test smells

Amjed Tahir
*School of Engineering and Advanced Technology Massey University Palmerston North, New Zealand a.tahir@massey.ac.nz*

Steve Counsell
*Department of Computer Science Brunel University London, United Kingdom steve.counsell@brunel.ac.uk*

Stephen G. MacDonell
*Department of Information Science University of Otago Dunedin, New Zealand stephen.macdonell@otago.ac.nz*

## Abstract

*While a substantial body of prior research has investigated the form and nature of production code, comparatively little attention has examined characteristics of test code, and, in particular, test smells in that code. In this paper, we explore the relationship between production code properties (at the class level) and a set of test smells, in five open source systems. Specifically, we examine whether complexity properties of a production class can be used as predictors of the presence of test smells in the associated unit test. Our results, derived from the analysis of 975 production class-unit test pairs, show that the Cyclomatic Complexity (CC) and Weighted Methods per Class (WMC) of production classes are strong indicators of the presence of smells in their associated unit tests. The Lack of Cohesion of Methods in a production class (LCOM) also appears to be a good indicator of the presence of test smells. Perhaps more importantly, all three metrics appear to be good indicators of particular test smells, especially Eager Test and Duplicated Code. The Depth of the Inheritance Tree (DIT), on the other hand, was not found to be significantly related to the incidence of test smells. The results have important implications for large-scale software development, particularly in a context where organizations are increasingly using, adopting or adapting open source code as part of their development strategy and need to ensure that classes and methods are kept as simple as possible.*

**Keywords:** test smells, OO metrics, software testing, empirical study

## 1. INTRODUCTION

A large number of empirical studies have explored the features and characteristics of production code and, particularly in recent years, those of open source production code. The increasingly automated collection of software metrics data has enabled us to capture many essential elements of that production code. By comparison, far less attention has been directed towards understanding the role that test code plays in the development and evolution of systems [1, 2]. Too often, test code has been treated as the 'poor relation' to production code, yet anecdotally, poorly written and maintained test code has the potential for a greater volume than corresponding production code. This stance is likely to have been exacerbated by the use of Agile methods, where great emphasis is placed on the rapid delivery of client-facing 'working software'. In fact, test code needs to be seen as an integral part of the 'working software', with similar 'status' as that accorded to production code.

Where large suites of system tests are run against production code, test code inefficiencies, unnecessary duplication, excessive and potentially harmful coupling, defects in the test suite and redundancy are often key and undesirable features. In spite of this, we still know very little about how system test suites evolve, how dependencies between tests reflect those in production code and how to more effectively manage growth in test suites. There is also increasing pressure to improve delivery times while maintaining test and production quality (i.e., the economics of testing); key to improvements is the streamlining and optimization of each. There are clear quantifiable benefits to improving test processes in terms of the time that could be devoted to other aspects of development were these costs re-directed (i.e., its opportunity cost). In short, the motivation for further empirical investigations is clear.

Studying the particular relationship between production code and test smells can assist early detection of those smells. Source code metrics are easier and faster to collect compared to test smell detection, as there is a large set of automated tools that are available to collect source code metrics, including those for complexity. In contrast, there are very few test smell detection and refactoring tools. Moreover, the currently available tools lack consistency and typically require manual verification (explained further in Section 3.2). In contrast, most complexity metrics can be collected during early stages of development (e.g., at the design stage).

In this paper we explore the incidence of test smells and use four software metrics to inform our analysis; all four are surrogates for complexity and measure internal complexity [3], size, cohesion and inheritance [4]. We use five open source systems comprising 975 unit tests to address two research questions related to test smells: firstly, whether the number of test smell types in a unit test was related to the complexity properties of its corresponding production class; secondly, whether complexity-related properties of production classes varied among unit tests exhibiting different kinds of test smell types. The remainder of the paper is structured as follows: Section 2 describes our study background and related work. Section 3 explains our empirical study design and outlines our research questions. We present our results in Section 4 followed by a discussion of our findings. We consider a number of threats to the validity of the study in Section 5. Finally, we provide concluding remarks and directions for future work in Section 6.

## 2. BACKGROUND AND RELATED WORK

Test smells are design flaws or issues that appear in test code (and typically in unit tests). The term was coined by van Deursen et al. [5] to refer the types of code smells that particularly affect unit tests, following the definition by Fowler et al. [6] of twenty-two code smells. Code smells are symptoms of possible refactoring opportunities [6], whose purpose is to improve code readability, maintainability and understandability. In other words, to remove a code smell, a set of refactorings is applied. Examples of smells include an excessively large class (Large Class [6]) or excessively long method (Long Method [6]).

Code smells have been shown anecdotally and in limited literature to have a negative impact on software quality characteristics [7-9]. In a similar way, test smells have been shown to negatively impact test readability and comprehension during maintenance activities [10]. Test code should therefore be checked on an ongoing basis for potential smells and, when found, appropriate refactoring mechanisms put in place to remove (or at least reduce) those smells. Several specific test smell refactoring approaches have been suggested in the literature [5, 11] but there is only limited empirical evidence to support conjectures about test code quality. Several studies have investigated code smell distribution in software systems and their potential impact on software quality attributes [7, 8, 12, 13]. However, less effort has been directed towards test smells by comparison. Most test smell studies focus on either defining new smells [5, 14], or designing smell detection techniques and tools [15, 16].

To the best of our knowledge, only three empirical studies have assessed the presence of test smells and/or their impact on software artefacts. Bavota et al. [10] studied the presence of eleven tests smells (as defined by van Deursen et al. [5]) in both open source and industrial Java systems. They found that test smells were widely distributed in both open source and industrial systems, with almost 86% of the examined unit tests containing at least one test smell. The study also examined the impact of test smells on test comprehension during typical maintenance tasks through a controlled experiment using students and practitioners. The presence of test smells, generally, had a negative impact on developers' comprehension during the defined maintenance activities. Greiler et al. [17] investigated how six fixture-related test smells evolved over time in open source projects. The study reported that test fixture smells did not continually increase over time (from one release to another), even when the system's complexity increased. In recent work, we studied how test smells were influenced by several properties of production code and found that increasing levels of size and Cyclomatic Complexity in code can have a negative impact on the presence of test smells [18]. In addition, there are particular code smells that can be seen as 'signs' for the presence of test smells; some test and code smell types co-occur in production classes and their associated unit tests [18]. In another recent study, Palomba et al. [19] inspected the presence of test smells in automatically generated unit tests by EvoSuite – a popular automatic test suite generation tool for Java. The results showed high presence of Assertion Roulette, Duplicated Code and Eager Test smells (see Table 1 for definitions) in automatically generated unit tests. The authors also report that there are several characteristics of EvoSuite that contribute to the introduction of test smells when generating test suites.

In this study, we perform a fine-grained analysis to explore the relationship between properties of production classes and the presence of test smells in the unit tests developed to test those classes. The goal is to examine whether class attributes in OO systems have an impact on the presence of test smells in associated unit tests.

## 3. EXPERIMENTAL DESIGN

In this work we focus on four OO complexity-related aspects: *internal complexity, weighted methods complexity, inheritance and cohesion*. We measure class internal complexity using McCabe's Cyclomatic Complexity (CC) [3], *weighted methods complexity* via the Weighted Methods per Class (WMC) metric, class *inheritance* using Depth of the Inheritance Tree (DIT) and cohesion using the Lack of Cohesion of Methods (LCOM) metric. All four metrics represent different - albeit related - aspects of class complexity. WMC, DIT and LCOM are all part of the Chidamber and Kemerer (C&K) metrics suite [4]. Previous studies have found strong associations between complexity metrics and class fault-proneness [20]. Generally, a software design should aim to minimize the values of CC, WMC, DIT and LCOM for individual classes.

**CC:** is a measure of the number of independent paths of each method within a class. The CC of a class is the average CC of all methods within that class. A method with more control flow statements (e.g., conditional statements) will essentially have more paths. Higher numbers of independent paths in a class typically require more developer effort in order to read, understand and test (close to) all possible independent paths. Increases in the value of CC (particularly those of value higher than 10) is associated with an increased number of faults [21]. This metric is known for its popularity among practitioners, mainly due to its ease of collection [22].

**WMC:** is the count of the number of methods, weighted based on their complexity values. The larger the number of methods in a class, the greater the potential impact on children, since children will inherit all methods defined in the class. Classes with large numbers of methods are likely to be more application-specific, limiting the possibility of reuse. The original definition of WMC included a subjective measure of the complexity of each method and the WMC of a class would be summation of those individual complexities; however, C&K later re- defined the complexity of a method to become simply '1'. In other words, the later definition of WMC is equivalent to simply the number of methods in a class.

**DIT:** is the maximum length from the node to the root of the inheritance tree. DIT is a measure of how many ancestor classes can potentially affect a class. The deeper a class is in the hierarchy, the greater the number of methods it is likely to inherit, making it, in theory, more complex [23]. In Java, the root of the tree is Object, and is considered to be at level zero. All Java classes that inherit directly from Object are therefore positioned at DIT level 1.

**LCOM:** measures the overlap of class fields in the methods of a class. If every method uses every field, then that class is maximally cohesive (LCOM value 0). If the use of fields in methods is completely disjoint (i.e., each method uses a different field or set of fields) then that class is minimally cohesive (LCOM of 1). Method cohesiveness within a class is highly desirable, since it promotes encapsulation. Lack of cohesion implies classes should probably be split into two or more classes/sub-classes. Low cohesion increases complexity, thereby increasing the likelihood of human error during the development process and faults being injected in the code.

Having a means to reliably link production classes to their unit tests is critical to this work. In this regard we use two well-established test- to-code traceability techniques: unit tests *Naming Convention and Static Call Graph* [24]. First, we use the *Naming Convention* technique, which reflects the widely suggested unit testing practice - especially within the xUnit framework - that a unit test should be named after the corresponding class that it tests, by adding "Test" to the original class name. For example, the unit test for class 'foo' should be 'fooTest'. We then use the *Static Call Graph* technique, which inspects method invocations in the test case. Both techniques are applied manually. The effectiveness of the *Naming Convention* technique is reliant on developers' efforts in conforming to the recommended coding standard, whereas the *Static Call Graph* technique reveals direct references to production classes in the unit tests. As explained by van Rompaey and Demeyer [24], test-to-code traceability strategies such as *Naming Convention* result in high precision and recall, but they also depend highly on the testing strategy followed. Other strategies such as Static Call Graph and Last Call before Assertion have high applicability but score low in accuracy. It has therefore been recommended that a combination of strategies be used [24], a recommendation we have followed here.

TABLE 1: LIST OF TEST SMELLS AND THE TOOLS USED TO COLLECT THEM

| | Test Smell | Description | Tool |
|---|---|---|---|
| 1 | Assertion-free | A test case without assertion.[1] | PMD |
| 2 | Assertion Roulette | Test method having more than one assertion. | PMD |
| 3 | Sensitive Equality | The toString method is used in assert statements. | [10] |
| 4 | Mystery Guest | A test case that uses external resources. | [10] |
| 5 | Indirect Test | A test allocates resources also used by others. | [10] |
| 6 | General Fixture | A test fixture is too general and the test methods access only part of it. | [10] |
| 7 | Duplicated Code | Sets of test commands that contain the same invocation and data access sequence. | CodePro |
| 8 | Eager Test | A unit test has at least one method that uses more than one method of the tested class. | [10] |
| 9 | Lazy Test | Several test methods check a method of the tested class using the same fixture. | [10] |

Table 1 lists the set of nine test smells considered in this study, together with the tool that was used to collect each one. We selected these smells because: 1) they are well-defined in the literature, 2) they have been considered in similar previous studies, and 3) there were tools available to detect these smells.

## 3.1. Research questions and hypotheses
In this paper we investigate the following research questions and hypotheses:

**RQ1** Is the number of test smell types in a unit test related to the complexity properties of the corresponding production class?

$H1_0$ There is no significant relationship between the number of test smell types present in a unit test and complexity-related properties of the corresponding production class.

$H1_1$ There is a significant relationship between the number of test smell types present in a unit test and complexity-related properties of the corresponding production class.

**RQ2** Do complexity-related properties of production classes vary for associated unit tests that exhibit different test smell types?

$H2_0$ There is no significant difference between the complexity- related properties of production classes where their associated unit test contains different test smells types.

$H2_1$: There is a significant difference between the complexity- related properties of production classes where their associated unit test contains different test smells types.

---

[1] Assertion-free can be seen as a test defect (given that it does not perform any testing) or a test smell (a missing test).

TABLE 2: SYSTEM CHARACTERISTICS

| System | Version | Size (KLOC) | #Classes | #Unit tests | #Analyzed unit tests | Release date | Age Years | # Contributors (all releases) | Class coverage level | URL |
|---|---|---|---|---|---|---|---|---|---|---|
| JFreeChart | v. 1.0.17 | 140.6 | 669 | 366 | 359 | 24-11-2013 | 15 | 8 | 83% | http://www.jfree.org/jfreechart |
| JabRef | v. 2.9.2 | 106.4 | 1150 | 127 | 75 | 12-01-2013 | 12 | 38 | 47% | http://jabref.sourceforge.net |
| Apache Commons Lang | v. 3.3.2 | 63.6 | 132 | 142 | 124 | 09-04-2014 | 13 | 15 | 100% | http://commons.apache.org/proper/commons-lang |
| Dependency Finder (DF) | v. 1.2.1-beta4 | 58 | 450 | 280 | 220 | 29-11-2010 | 12 | 2 | 61% | http://depfind.sourceforge.net |
| MOEA | v. 1.17 | 42 | 407 | 209 | 197 | 14-11-2012 | 4 | 1 | 87% | http://www.moeaframework.org |
| Total | ----- | 410.6 | 2808 | 1124 | 975 | ----- | ----- | ----- | ----- | ----- |

## 3.2. Empirical analysis

We used five open source systems in our analysis, as follows: JFreeChart (a Java chart and graphics library), JabRef (a bibliography management tool that supports BibTeX files), Apache Commons Lang (a set of helper utilities for the java.lang API), Dependency Finder (a tool that extracts dependencies and dependency metrics of complied Java code) and MOEA (a framework that supports development of multi-objective evolutionary and optimization algorithms). The general characteristics of the five selected systems are shown in Table 2. The lowest class coverage was 61% (in Dependency Finder), while the highest was 100% (for Commons Lang). The systems varied in age – JFreeChart was 15 years old, MOEA just 4 years old. In total, we analyzed 975 unit tests and their corresponding production classes, finding a total of 1511 smell instances in the unit tests. To assure the accuracy of our results we verified the detected smells qualitatively through a manual inspection process conducted by the first author. All false-positive smell instances were removed from the analysis, while all false-negative smell instances were included in our dataset[2]. We also cross-validated some of our results with those obtained by Bavota et al. [10] for one of the systems (i.e., Dependency Finder) considered in their work, using the authors' publicly available data. We did this for 118 of the 220 unit tests for the following smells: Sensitive Equality, Mystery Guest, Indirect Test, General Fixture, Eager Test and Lazy Test. These smells found 10 false-positive smell instances. All false positive results were re-checked and revalidated manually, before they were removed. Due to these checks we are confident in the veracity of the smell detection process followed in this work.

A number of statistical procedures were used in order to analyze the collected data. We first used the Shapiro-Wilk test to check whether the various metrics' data distributions adhered to the characteristics and assumptions of a normal distribution. We found that the data were not normally distributed (in common with much software-related data). We therefore used the non-parametric Spearman's rho (ρ) rank correlation coefficient to test for statistical association between independent variable pairs (i.e., between individual metrics and the number of test smell types) (RQ1). We used Cohen's classification [25] to interpret the degree of association (measured using Spearman's rho, ρ) between variables: a low association is assumed when $0 <$

$p < 0.3$, medium when $0.3 \leq p < 0.5$ and high when $p \geq 0.5$. This interpretation also applies to negative correlations (where the association is inverse rather than direct).

To compare differences in distribution between two independent groups (i.e., between individual metrics and individual test smells), we used the non-parametric Mann–Whitney U test (RQ2). In the analyses reported in this paper, we interpreted there to be a significant relationship between two variables if there was statistically significant evidence of such a relationship in at least three of the five systems examined.

TABLE 3: NO. OF UNIT TESTS CONTAINING DIFFERENT TEST SMELL TYPES

| Test Smell | JFree Chart | JabRef | Commons Lang | DF | MOEA | Total |
|---|---|---|---|---|---|---|
| Assertion-free | 9 | 5 | 31 | 15 | 8 | 63 |
| Assertion Roulette | 347 | 51 | 85 | 172 | 140 | 744 |
| Sensitive Equality | 6 | 5 | 58 | 11 | 2 | 77 |
| Mystery Guest | 0 | 10 | 5 | 29 | 23 | 57 |
| Indirect Test | 0 | 1 | 2 | 22 | 3 | 27 |
| General Fixture | 28 | 9 | 16 | 112 | 27 | 183 |
| Duplicated Code | 59 | 17 | 36 | 85 | 66 | 246 |
| Eager Test | 33 | 8 | 14 | 30 | 19 | 96 |
| Lazy Test | 9 | 2 | 5 | 0 | 4 | 18 |
| **Total** | 491 | 108 | 252 | 476 | 292 | 1511 |
| **Mean** | 54.6 | 12 | 28.0 | 52.9 | 32.4 | 167.9 |
| **Std Dev** | 111.3 | 15.4 | 28.1 | 57.9 | 44.9 | 228.4 |

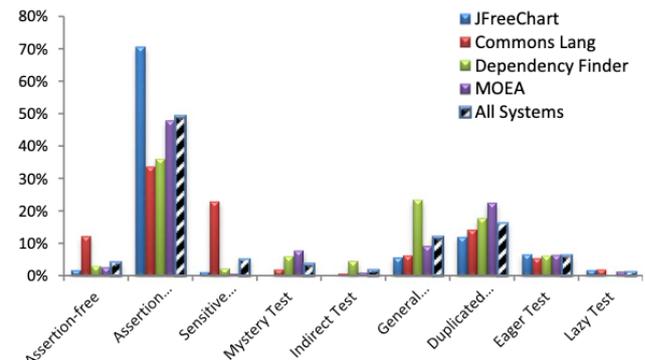

Figure 1: Distribution of test smell types in all five systems

## 4. RESULTS AND DISCUSSION

In this section we report the results of the analyses conducted to answer the two research questions stated in Section 3.1, in each case accompanied by a discussion of the presented results. We first show the distributions of smells across all five systems (Table 3), and the relative proportions of unit tests that contain at least one smell across all five systems are shown in Figure 1. In considering the occurrence of each individual test smell, it is important to highlight that these smells are measured in terms of their presence (or not) in a unit test. For example, a unit test can either have *Assertion Roulette* (true) or not (false), irrespective of the number of times the smell was detected in the unit test. Therefore, with regard to the incidence of test smells in a unit test, that test can have a value between zero (i.e., minimum value, that is, no smells present) and nine (i.e., maximum value, contains all nine smell types). We refer to this as the "number of test smell types".

---

[2] In total, we found 6 false-positive instances in JabRef and 10 instances in Dependency Finder. This corresponds to approximately 9% of the total number of smells in these two systems.

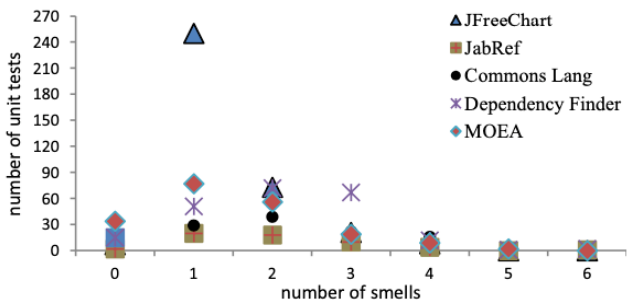

Figure 2: Distribution of test smells in all five systems

As shown in Table 3, and visually in Figure 1, Indirect Test and Lazy Test smells rarely occurred, so these have been excluded from any further analysis. Indirect Test comprises only around 2%, and Lazy Test 1%, of the total number of test smells in these systems. The comparative absence of these smells was also a feature of the empirical analysis reported by Bavota et al. [10], where just 2% of the overall set of smells fell into the category of Lazy Test. The same study concurred in terms of the most commonly encountered test smell; in the study by Bavota et al. [10], the most frequently occurring test smell type was Assertion Roulette (at 55%); we note that in [10], 27 systems were studied, overlapping with two systems considered in this study (i.e., JabRef and Dependency Finder).

Figure 2 depicts the distribution of test smells across all five systems. As is evident in the graph, the majority of unit tests contain at least one test smell. It is also clear that the vast majority of unit tests contain between 1 and 3 smell types (86% of the unit tests). Very few unit tests are smell-free (~ 8%). The number of unit tests that contain more than 3 smell types decreases significantly in four systems. The only exception here is JFreeChart, where the majority of unit tests (70%) contain only one test smell type (and most likely it is *Assertion Roulette*, due to the high spread of this smell in this system, as per Figure 1). A very limited number of unit tests contain more than 4 smell types (less than 1% of the number of unit tests).

In Table 4 we report descriptive statistics for the four metrics we collected from the 975 production classes (of all five systems) associated with the unit tests considered in this analysis. In the results reported in the remainder of this section, we interpreted there to be a significant relationship between two variables if there was statistically significant evidence of such a relationship in at least three of the five examined systems.

TABLE 4: DESCRIPTIVE STATISTICS OF THE METRICS DATA

| Systems | | Min | Max | Mean | Std Dev |
|---|---|---|---|---|---|
| **JFreeChart** | CC | 0 | 21 | 2.35 | 1.53 |
| | WMC | 0 | 666 | 54.08 | 70.51 |
| | DIT | 1 | 6 | 3.23 | 1.17 |
| | LCOM | 0 | 0.97 | 0.37 | 0.33 |
| **JabRef** | CC | 1 | 26 | 5.29 | 5.36 |
| | WMC | 1 | 520 | 57.35 | 104.96 |
| | DIT | 1 | 3 | 2.10 | 0.40 |
| | LCOM | 0 | 0.85 | 0.13 | 0.25 |
| **Commons Lang** | CC | 1 | 10 | 2.39 | 1.98 |
| | WMC | 1 | 816 | 97.48 | 179.49 |
| | DIT | 0 | 5 | 0.69 | 1.05 |
| | LCOM | 0 | 1 | 0.20 | 0.34 |
| **Dependency Finder** | CC | 0 | 20 | 1.93 | 1.94 |
| | WMC | 0 | 163 | 23.55 | 31.14 |
| | DIT | 0 | 4 | 1.52 | 1.22 |
| | LCOM | 0 | 1 | 0.32 | 0.34 |
| **MOEA** | CC | 1 | 38 | 2.39 | 2.96 |
| | WMC | 1 | 153 | 16.40 | 19.35 |
| | DIT | 1 | 5 | 2.52 | 0.65 |
| | LCOM | 0 | 0.93 | 0.15 | 0.27 |

TABLE 5: SPEARMAN'S  P CORRELATION BETWEEN THE NUMBER OF TEST SMELL TYPES AND THE FOUR COMPLEXITY-RELATED METRICS OF THE CORRESPONDING PRODUCTION CLASS (all significant values (i.e., p ≤ 0.05) are marked with an asterisk (*) and all *medium* and *high* correlations are shown in **bold**)

| Metrics | | Number of Test Smell Types | | | | |
|---|---|---|---|---|---|---|
| | | **JFreeChart** | **JabRef** | **Commons Lang** | **Dependency Finder** | **MOEA** |
| CC | ρ | .24* | .22* | **.56*** | **.35*** | **.30*** |
| | p | .00 | .05 | **.00** | **.00** | **.00** |
| WMC | ρ | **.51*** | **.58*** | **.52*** | **.38*** | **.33*** |
| | p | **.00** | **.00** | **.00** | **.00** | **.00** |
| DIT | ρ | .19* | .06 | -.25* | -.20* | .46 |
| | p | .00 | .34 | .00 | .00 | .91 |
| LCOM | ρ | .24* | **.47*** | .16* | **.30*** | **.26*** |
| | p | .00 | **.00** | .04 | **.00** | **.00** |

For RQ1, the results of the correlation analyses between all four metrics and the number of test smell types in a unit test are reported in Table 5. The results show that CC, WMC and LCOM are all significantly correlated with the number of test smell types across all five examined systems, although the correlation between LCOM and the number of test smell types is generally weaker than for the other two metrics. There is a high correlation between CC and the number of test smell types in Commons Lang, a medium correlation in Dependency Finder and MOEA, and a low correlation in JFreeChart and JabRef. WMC is also significantly correlated with the number of test smell types - with high correlation in JFreeChart, JabRef and Commons Lang, and medium correlation in Dependency Finder and MOEA. There is also a significant correlation (albeit weaker than CC and WMC) between LCOM and the number of test smell types. On the other hand, DIT shows mixed results; JFreeChart show a low significant correlation with the number of test smell types, two of the five systems (i.e., Commons Lang and Dependency Finder) show negative correlations with the number of test smell types (i.e., an increase in the depth of the inheritance tree is mirrored by a decrease in the presence of test smells in the associated unit test), while two other systems show no correlations (i.e., JabRef and MOEA).

TABLE 6: MANN-WHITNEY U TEST RESULTS FOR THE RELATIONSHIPS BETWEEN THREE COMPLEXITY METRICS AND INDIVIDUAL TEST SMELLS (all significant values (i.e., p ≤ 0.05) are marked with an asterisk (*). Results that are significant for *three or more* systems are highlighted)

| Smells | Metrics | JFreeChart | JabRef | Commons Lang | Dependency Finder | MOEA |
|---|---|---|---|---|---|---|
| **Assertion-free** | CC | .13 | .45 | .96 | .83 | .40 |
| | WMC | .00* | .55 | .16 | .55 | .02* |
| | DIT | .27 | .49 | .35 | .90 | .23 |
| | LCOM | .17 | .97 | .09 | .47 | .11 |
| **Assertion Roulette** | CC | .16 | .71 | .21 | .00* | .31 |
| | WMC | .14 | .96 | .00* | .03* | .74 |
| | DIT | .10 | .42 | .01* | .30 | .21 |
| | LCOM | .21 | .80 | .08 | .01* | .69 |
| **Sensitive Equality** | CC | .08 | .04* | .09 | .29 | .99 |
| | WMC | .64 | .02* | .28 | .01* | .44 |
| | DIT | .89 | .01* | .10 | .11 | .20 |
| | LCOM | .59 | .40 | .01* | .02* | .73 |
| **Mystery Guest** | CC | ----- | .68 | .50 | ----- | .01* |
| | WMC | ----- | .12 | .50 | .50 | .00* |
| | DIT | ----- | .22 | .34 | .30 | .63 |
| | LCOM | ----- | .11 | .15 | .42 | .01* |
| **General Fixture** | CC | .19 | .61 | .18 | .19 | .23 |
| | WMC | .00* | .11 | .18 | .22 | .00* |
| | DIT | .05* | .29 | .55 | .04* | .26 |
| | **LCOM** | .64 | .02* | .41 | .00* | .00* |
| **Eager Test** | **CC** | .00* | .68 | .01* | .00* | .16 |
| | **WMC** | .00* | .01* | .00* | .03* | .00* |
| | DIT | .01* | .23 | .07 | .31 | .57 |
| | **LCOM** | .00* | .07 | .39 | .01* | .00* |
| **Duplicated Code** | **CC** | .01* | .00* | .02* | .01* | .00* |
| | **WMC** | .00* | .00* | .00* | .00* | .17 |
| | DIT | .06 | .95 | .00* | .01* | .56 |
| | **LCOM** | .01* | .28 | .78 | .00* | .04* |

The positive correlation between WMC and the number of test smell types might well be expected; prior evidence has indicated that the larger an artefact, the greater the presence of faults, for example. The same would likely be true for the presence of different test smells, as potential precursors of faults. The correlation values for DIT, on the other hand, show a strong negative, significant relationship for two of the systems (Commons Lang and Dependency Finder). A number of interpretations could be made for this finding. Firstly, in accordance with the principles of inheritance, the amount of code at lower levels of inheritance might preclude the emergence of large amounts of code and hence smells at these levels. Secondly, the difference between the results for the pairs of systems may be due to the fact that each pair used inheritance depth more (or less) extensively. In other words, the nature of the respective applications might lend themselves to the use of inheritance at shallower levels of inheritance and with larger classes at each level as a direct consequence. Table 4 supports this thesis in showing that the DIT values for Commons Lang and Dependency Finder are the smallest of the five systems. The hierarchies of these two systems are more compressed vis-à-vis JFreeChart, JabRef and MOEA. It is worth noting that DIT has often been shown, in most systems, to have lots of tied values for it towards the lower end of the scale – and so it does not discriminate well. That is, classes can have the same (low) value for DIT but wildly varying values for the variables of interest – in this case, the incidence of test smells.

```
protected void setUp() throws Exception {
    parserFactory = SAXParserFactory.newInstance();

    saxParser = parserFactory.newSAXParser();

    be = new BibtexEntry(Util.createNeutralId(),
        BibtexEntryType.getType("article"));

    handler = new OAI2Handler(be);
    }
```

Figure 3: An example of General Fixture smell from JabRef.

In general, we observed that the number of test smell types in a unit test increases as the complexity of the associated production class increases. Several relationships between OO metrics (such as coupling, cohesion and Cyclomatic Complexity) and unit test size have been found in previous research [26]. In our own prior work we found evidence of a significant association between several dynamic complexity metrics (such as dynamic coupling and execution frequency) and the size of unit tests [27]. Given this broad body of evidence, we find the correlation between class complexity and the number of test smell types in the associated unit tests to be plausible. Developers may attempt to test a complex class by writing a large unit test, which in turn increases the chances of introducing test smells into the unit test. We also need to consider the role that copy and paste operations across unit tests might contribute to duplicate and redundant code in tests. Code 'bloat' [6] in tests can arise (just as it can in production code) through unnecessary reuse of test code.

It is important to note that herein we did not investigate the causation (i.e., direction) of these relationships. It is therefore not possible to say if any of the variables considered *completely* and directly influences the others (i.e., whether production code properties influenced the test code, or *vice versa*). Determination of the relationship direction in this case will depend highly on the testing strategy followed (i.e., test-first or test-last). That said, it is also important to note that a recent empirical study by Beller et al. [28] noted that the majority of open source projects and developers/contributors follow a test-last approach, i.e., test code is developed after the production code. It was found that test-driven development (TDD) was not widely practiced – in only 12% of the projects that claimed to follow TDD did the developers actually use that strategy. In light of these findings, we could cautiously assert that the direction of the relationships identified in this paper is likely to follow a [production class → unit test] direction. Therefore, we may *cautiously* assume that production code properties *impact* the presence of test smells in test code. This assumption applies to both RQ1 and RQ2.

In revisiting our hypotheses for RQ1, we reject H1₀ and accept H1₁: there is a significant positive correlation between the number of test smell types present in a unit test and complexity-related properties of the corresponding production class.

The results of our investigation into the relationship between the four metrics and *individual* test smells (using the Mann-Whitney U test) are depicted in Table 6. Significant results in three or more systems are highlighted. In RQ2, we study whether unit tests containing individual test smells have larger complexity metric values for their associated production classes than other classes. It is evident from Table 6 that only three of the individual test smells studied (i.e., *General Fixture*, *Eager Test and Duplicated Code*) appear to be significantly related with at least one of the complexity-related metrics. Note that DIT did not appear to be significantly associated with any of the test smells in any of the five systems.

As shown in Table 6, unit tests that contain one or more General Fixture smells are significantly correlated with the LCOM of their associated production classes in JabRef, Dependency Finder and MOEA. This association is not evident in JFreeChart or Commons Lang. An example of General Fixture from JabRef is shown in Figure 3. This result regarding General Fixture is in line with the findings of Greiler et al. [17], which explained how other test fixture related smells were often correlated with the number of test cases. The increase in the number of test cases in a unit test is somehow a reflection of the increase of the size and complexity of the associated production class – a large or complex class will require more test cases.

```
private ClassfileScanner scanner;

public void testOneFile() {
    String filename = TEST_FILENAME;
    assertTrue(filename + " missing", new File(filename).exists());
    scanner.load(Collections.singleton(filename));
    assertEquals("Number of files", 1, scanner.getNbFiles());
    assertEquals("Number of classes", 1, scanner.getNbClasses());
}
```

Figure 4: An example of Eager Test smell from Dependency Finder.

We also found that unit tests that contain the Eager Test smell are significantly associated with large CC and WMC values in production classes, compared to unit tests that do not contain this smell (i.e., the larger the complexity of a class, the higher the chance of introducing Eager Test in the associated unit test). Similarly, unit tests that contain the Eager Test smell are associated with a high value of LCOM in three of the five systems (with JabRef and Commons Lang being the exception). To test a complex class, with high values of CC and WMC metrics, a unit test might attempt to test multiple methods of the same production class, leading to possible inter-dependencies between methods in the same class. In doing so, a test might be required to invoke several other related methods connected to the target method under test, which can lead to the introduction of the Eager Test smell – a test has a method that uses more than one method of the tested class. We show an example of Eager Test from Dependency Finder in Figure 4.

Finally, unit tests exhibiting the Duplicated Code smell are associated with high production class values for WMC (except in MOEA) and LCOM (except in Commons Lang). These relationships have a number of possible explanations. When testing a complex class, a test case might be required to invoke several methods from the same class, and since methods of the same class can have a similar structure, this might lead to an increase in the likelihood of creating code duplication.

On the balance of evidence, we accept H2₁ for the relationship between the CC, WMC and LCOM metrics and the Eager Test and Duplicated Code smell types: there is a significant difference between the values of CC, WMC and LCOM of production classes where their associated unit tests contain/do not contain the Eager Test and Duplicated Code test smells. Also, we also accept H2₁ for the relationship between LCOM and General Fixture smell: there is a significant difference between the value of LCOM of production classes where their associated unit tests contain/do not contain the General Fixture test smell.

## 5. THREATS TO VALIDITY

There are a number of threats that could affect the utility of the results presented in this paper. As explained in Section 3.2, all the metric and test smell data used in the analyses were initially collected using automated tools. Tool adequacy in correctly capturing this data could be challenged. To address this threat, we selected tools that have been used and validated in previous similar studies. Where possible, metrics data collected by one tool was cross-validated by those obtained from a similar tool3 (for example, metrics data were collected using CodePro Analytix and were validated using the Eclipse Metrics plugin). The tools used are also recognized as industry strength, limiting to some extent the possibility of false positives and negatives in the collected data. We also used three different automated tools to locate and identify test smells. This may increase the risk of including false-positive or false-negative smell instances. To improve the accuracy of our detection mechanism, we selected open-source tools that have been successfully used and evaluated in similar previous empirical studies. For most of the detected smells, we used an academic tool that has been successfully used in previous research [10], in addition to two well-known and widely used open-source source code analyzers (i.e., PMD and CodePro). To assure the accuracy of the results, we verified the detection of smells through a manual inspection process conducted by the first author, and we cross-validated some of our results with those obtained for the same systems considered in prior research [10] using those authors' publicly available data (details of this cross-validation process are provided in Section 3.2).

Another possible threat to the validity of the results presented in this paper relates to the identification and selection of production classes. Only production classes that have corresponding unit tests are included, which may lead to a form of selection bias. Since only one of the examined systems (i.e., Commons Lang) has unit tests available for all production classes (i.e., they provide complete class coverage), classes that are extremely difficult to test, or are considered too simple, might have no associated unit test; hence, they would not be considered in our analyses. However, we intentionally selected systems that have adequate to high class coverage (with average class coverage of 76%). Note that, in an open-source context, it is quite rare to find a system with 100% class coverage.

One further threat is that we have used open source as opposed to proprietary (commercial) systems in the study. Analysis of the latter might provide insights that could not be gleaned from our study (and we plan this as future work). In this paper, we considered five open source systems of differing application domains. This could be criticized for a lack of external validity; however, the same criticism would apply had we studied systems all from a similar application domain. It is also difficult to determine an ideal number of systems to study and what that limit might be.

Finally, every effort was made to apply statistical techniques that were appropriate to non-normally distributed data (e.g., rank order correlation tests).

## 6. CONCLUSIONS AND FUTURE WORK

In this study we investigated the impact of several production code properties on the presence of test smells in the associated unit tests of five open source systems. We used four metrics that capture important OO properties of production classes as surrogates for class complexity, measuring internal complexity, weighted methods complexity, inheritance and cohesion properties. We measured these properties using Cyclomatic Complexity and three metrics from the C&K metrics suite. Our results from analyzing 975 classes and their associated unit tests show that the CC, WMC and LCOM values of production classes are significantly associated with the incidence of unit test smell types. DIT, on the other hand, does not appear to be significantly related to the spread of test smells. Specifically, CC and WMC can be used as a strong indicator of the presence of test smells in the associated unit test; the LCOM metric also appears to be a good indicator of smells. Also, two metrics appear to be a 'sign' of the presence of several individual test smells: unit tests containing the Eager Test and Duplicated Code test smells are mostly associated with classes with high LCOM and high WMC values. We believe that conforming to OO best practice (i.e., to reduce the values of WMC and LCOM) could help in decreasing the likelihood of introducing test smells in test code.

Two implications of our results are relevant to industry practice. Firstly, the study reinforces the need for developers to minimize the number of methods in classes and the size of those methods as a means of controlling complexity and understandability. Secondly, refactoring [6] has a key role in ensuring that this can be achieved, although we accept that developers have only limited time and resources for such refactoring activity and that this might be an inhibiting factor.

This study is part of our ongoing effort to understand test smells and their possible impact on software development, especially during maintenance tasks. We intend to continue studying the factors that contribute to the appearance of smells in test code. In the near future, we intend to investigate whether such smells have an impact on defect density in software; for example, does the presence of test smells in the unit test have anything to do with defects that go undetected? We are also interested to see if test smells evolve over releases/time (i.e., where and when test smells appear in a unit test) and how to use such information to manage/control the spread of test smells; little research currently exists on the evolution of code smells and their potential for evolving into code 'stenches'. Finally, it would be worthwhile exploring other software metrics than those used in this paper. One example might be based on coupling and its decomposition into incoming and outgoing class coupling (fan-in and fan-out). The relationships found here (and possibly for other similar production code metrics) could be used as predictors for tests smells. A prediction model can then be tested using class metric information to predict the presence of test smells in the associated unit tests. Such an approach could be then compared to the more traditional approach of using smell detection tools to detect test smells. Finally, replicating our study on commercial systems would likely also inform our overall understanding of test smells.

---

3 Though the use of multiple tools was not possible in many cases due to the lack of availability of such tools.

## ACKNOWLEDGMENT

The work by Steve Counsell was partly funded by the Engineering and Physical Sciences Research Council (EPSRC) of the UK under grant ref: EP/M024083/1